\documentclass[prd,nofootinbib,preprintnumbers,
  reprint,
  ]{revtex4-2}
\usepackage{amssymb,amsmath}
\usepackage{graphicx}
\usepackage[T1]{fontenc}
\usepackage{lmodern}

\usepackage[colorlinks=true
  ,urlcolor=blue
  ,citecolor=blue
  ,linkcolor=blue
]{hyperref}

\newcommand{\GeV}{\text{\,GeV}}
\newcommand{\eV}{\text{\,eV}}

\begin{document}

\preprint{MAN/HEP/2020/010}

\title{Heavy Light Inflaton and Dark Matter Production}

\author{Fedor Bezrukov}
\email{Fedor.Bezrukov@manchester.ac.uk}
\author{Abigail Keats}
\email{Abigail.Keats@postgrad.manchester.ac.uk}

\affiliation{Department of Physics and Astronomy,
  University of Manchester, Manchester, M13 9PL, United Kingdom}

\date{October 2020}

\begin{abstract}
We study the minimal extension of the SM by a scalar with quartic interaction serving as an inflaton. For the model where scale symmetry is broken only in the inflaton sector, the mass of the inflaton is constrained to be relatively low. Here, we analysed the previously omitted situation of the inflaton masses $m_{\chi}\gtrsim 250\text{ GeV}$.  Therefore, we provide a window of inflaton masses with viable inflationary properties that evade direct observational constraints, due to their small mixing with the Higgs sector. The addition of heavy neutral leptons with Majorana masses induced by the interaction with the inflaton allow for Cold Dark Matter in the model with masses  $O(1-10)\text{\,MeV}$.
\end{abstract}

\maketitle

\section{Introduction}

The observable Universe is homogeneous and isotropic, and almost completely flat. It is filled with matter and radiation and has an almost invariant spectrum of primordial density perturbations. These seemingly finely tuned initial conditions can be explained by the presence of an inflationary epoch prior to the Hot Big Bang  \cite{Starobinsky:1980te,Mukhanov:1981xt,Guth:1980zm,Linde:1981mu,Albrecht:1982wi}. Viable inflationary models should also provide a mechanism to initiate a reheating period post-inflation, during which the Standard Model (SM) particles are produced. A complete realistic model should also have a Dark Matter (DM) candidate and means to produce it, to give over $80\%$ of the total matter energy density today \cite{PhysRevD,Bertone_2018}.

This paper studies the extension of the SM by a scalar inflaton with quartic self-interaction, which was suggested in \cite{Shaposhnikov:2006xi}. With the addition of a small non-minimal coupling to gravity, this model provides inflationary predictions in agreement with the Cosmic Microwave Background (CMB) observations \cite{Kaiser:1994vs,Komatsu:1999mt,Bezrukov:2013fca}. We assume here, similar to \cite{Shaposhnikov:2006xi,Anisimov:2008qs,Bezrukov:2009yw,Bezrukov:2013fca,Bezrukov:2014nza}, that the scale symmetry in the scalar sector is broken only by the symmetry breaking massive term of the inflaton. Thus, the scalar provides symmetry breaking in the Higgs sector, as well as inflation. Notably, it can also give Majorana masses to uncharged heavy neutral leptons (HNLs), which in turn can be used as DM particles in the model \cite{Shaposhnikov:2006xi,Bezrukov:2014nza}.

In the previous works \cite{Shaposhnikov:2006xi,Anisimov:2008qs,Bezrukov:2009yw,Bezrukov:2013fca,Bezrukov:2014nza} the analysis was focused on the region of parameters with the inflaton mass below the Higgs boson mass. Here we focus on another case, where the inflaton is heavier and the channel of its direct decay to a pair of Higgs bosons is open, which allows for a more efficient reheating. Requiring reheating to be above electroweak temperatures then provides the upper bound on the inflaton mass. DM production is also studied for this parameter window. Here DM is produced from direct inflaton decays during the reheating process, leading to a non-thermal velocity distribution for DM.  We provide both a numerical study of the DM production using Boltzman equations and analytic results for limiting cases.

The paper first outlines the scalar quartic inflationary model with a non-minimal coupling, including cosmological constraints on the inflaton self-coupling given by the scalar density perturbation amplitude and limits on the tensor-to-scalar ratio. We will then describe the non-thermal inflaton distribution resulting from turbulent preheating, which provides the initial condition for the reheating study and DM production, and find the relevant inflaton decay widths. The first part of our analysis analytically estimates the DM mass and average momentum at the end of reheating, calculated in the limit of reheating temperatures much less/greater than the inflaton mass \cite{Bezrukov_2016,Shaposhnikov:2006xi}. Next, we solve the Boltzmann equations numerically across the entire inflaton parameter space; here the inflaton mass is constrained, for a given self-coupling, by kinematics of the decay and the electroweak symmetry breaking scale. The analytical and numerical results are comparable only at the extremities of the parameter space. Our final results conclude that heavy inflaton decay in the early universe produces MeV sterile neutrinos; with an average momentum over temperature at the end of reheating of $O(1-10)$, they are classified as Cold DM candidates.

\section{The Model}

\subsection{Inflationary Model}

The scalar potential of the model, following \cite{Shaposhnikov:2006xi,Anisimov:2008qs,Bezrukov:2009yw}, with $X$ being the inflaton field, and $\Phi$ the Higgs doublet, is
\begin{equation}
    \label{eq:1}
    V(X,\Phi)=-\frac{1}{2}m_{X}^{2}X^{2}+\frac{\beta}{4}X^{4}+\lambda(\Phi^{\dagger}\Phi-\frac{\alpha}{\lambda}X^{2})^{2} .
\end{equation}
We assumed that the only source of scale symmetry violation is due to the negative mass term in the inflaton sector.  Then, the negative quartic inflaton-Higgs coupling allows for the transfer of symmetry breaking into the SM sector\footnote{The domain wall problem can be avoided with the addition of a small cubic term, $\mu\chi^{3}$ with $\mu\lesssim \sqrt{\alpha^{3}/\lambda} v$, not alterating the reheating dynamics and the inflaton mass \cite{Anisimov:2008qs}.}.
During the  slow roll inflationary evolution, the field values converge to the attractor solution along the gradient of the potential. At inflation the quadratic term can be neglected, and this attractor is given by the line  $\sqrt{2}\Phi=\theta_{\text{inf}}X$ in the field space, where the angle $\theta_{\text{inf}}$ is given by:
\begin{align}
\label{eq:1.2}
    \theta_{\mathrm{\text{inf}}}&=\sqrt{\frac{2\alpha+\beta}{\lambda}},
\end{align}
in the limit of $\alpha,\beta\ll\lambda$. 
Slow roll inflation terminates when $X\sim O(M_{\text{P}})$, after which scale invariance of the model is broken in the inflaton sector, giving rise to the vacuum expectation values (VEVs) of the SM Higgs boson and the scalar field \cite{Bezrukov:2013fca,Anisimov:2008qs,Bezrukov:2009yw}:
\begin{equation}
    \label{eq:2}
    \langle \Phi\rangle=\frac{v}{\sqrt{2}} , \qquad
    \langle X\rangle=v\sqrt{\frac{\lambda}{2\alpha}} ;
\end{equation}
with $v=246\GeV$ \cite{PhysRevD}. The angle of rotation of the vacuum with respect to the gauge basis $(h,\chi)=(\sqrt{2}\Phi -v,X-\langle X\rangle)$ is given by
\begin{align}
    \label{eq:3}
    \theta_{\text{v}}=\frac{\sqrt{2}\langle \Phi\rangle}{\langle X\rangle}=\sqrt{\frac{2\alpha}{\lambda}}.
\end{align}
Spontaneous symmetry breaking gives rise to the following masses of the excitations on top of the vacuum \cite{Bezrukov:2013fca,Anisimov:2008qs,Bezrukov:2009yw}:
\begin{equation}
    \label{eq:4}
    m_{\text{h}}=\sqrt{2\lambda}v , \qquad
    m_{\chi}=m_{\text{h}}\sqrt{\frac{\beta}{2\alpha}} ,
\end{equation}
which are found in a basis rotated by angle \cite{Bezrukov:2013fca,Anisimov:2008qs,Bezrukov:2009yw}
\begin{equation}
    \label{eq:5}
    \theta_{\text{m}}=\theta_{\text{v}} \ \frac{2\alpha}{2\alpha-\beta}
\end{equation}
with respect to the gauge basis\footnote{$\theta_{m}$ reduces to the angle given in \cite{Bezrukov:2013fca,Anisimov:2008qs,Bezrukov:2009yw} in the limiting case of light inflaton, where $2\alpha\gg\beta$.}. Measurement of $m_{\text{h}}=125\GeV$ \cite{PhysRevD} constrains the SM Higgs boson self coupling $\lambda\simeq0.1$. 

The quartic self-coupling term, $\beta$, is constrained by ensuring the model is consistent with the CMB measurement of the primordial scalar density perturbation amplitude \cite{infltheory}. Additionally, to put the model within the tensor-to-scalar ratio limit, $r<0.13$ \cite{2020}, a non-minimal coupling of the scalar field to gravity, $\xi X^{2}R/2$, is required \cite{Kaiser:1994vs,Komatsu:1999mt,Bezrukov:2013fca}. For a coupling in the range $O(10^{-5})\leqslant\xi< 1$, $\beta$ has the following limits \cite{Bezrukov:2013fca,Anisimov:2008qs}:  
\begin{align}
    \label{eq:7}
    O(10^{-13})&\leqslant\beta\leqslant O(10^{-9}) .
\end{align}
We limit our analysis with $\xi\lesssim1$, as larger $\xi$ would introduce a new energy scale $M_P/\xi$ below the $M_{\text{P}}$ scale.

\subsection{Preheating and Reheating}

At the end of inflation, the total energy density of the inflaton resides in the homogeneous oscillations of the inflaton field. Preheating for heavy inflaton, where $\beta>8\alpha$, proceeds slightly differently than for the light inflaton case, due to the misalignment of the inflationary attractor (\ref{eq:1.2}) with the position of the vacuum (\ref{eq:3}) in the field space. Thus, even for the study of the background dynamics, some inflationary energy is deposited in the oscillations of the Higgs field on top of the vacuum. We can approximate the magnitude of this effect by evaluating the ratio of the inflaton to Higgs fields' energy densities (for simplicity we take $\alpha\ll\beta$, so the Higgs field in the vacuum (\ref{eq:3}) is negligible compared to its value along the inflationary attractor \eqref{eq:1.2}):
\begin{align}
    \frac{\rho_{X}}{\rho_{\Phi}}\sim \frac{\beta X^{4}}{\lambda \Phi^{4}}\sim O\Bigg(\frac{\lambda }{\beta}\Bigg)\sim O(10^{8}-10^{12}) .
\end{align}
Therefore, this mechanism does not transfer noticeable energy into the Higgs-like direction during the early stages of preheating.

The preheating stage continues with energy transfer from oscillations of the zero mode into excitations of the Higgs and inflaton fields, in a way similar to the light inflaton case \cite{Anisimov:2008qs}.  Parametric resonance then quickly takes effect in exponentially exciting inflaton particles to occupy a highly non-thermal infra-red distribution function.  At the same time, the Higgs particles promptly re-scatter from their resonance bands due to their large self-coupling, $\lambda$, thereby preventing parametric enhancement from taking effect \cite{reh,therm}. Rescattering of inflatons from their resonance bands becomes significant once roughly half of the energy of the inflaton condensate has been transferred, after which the inflaton enters a phase of free-turbulence. During this period the inflaton distribution evolves self-similarly towards thermalisation, and so the following ansatz is used \cite{turb,therm}: 
\begin{align}
    f_{\chi}(k,\tau)=\tau^{-q}f_{\chi,0}(k\tau^{-p});
\end{align}
$\tau=t/t_{0}$ is a dimensionless time scale,  where $t_{0}$ is some arbitrarily late time, and $k$ is conformal momenta. The exponents, derived numerically using lattice simulations, are: $p\approx 1/5$ and $q\approx3.5p$ \cite{turb,therm}. The momentum distribution follows a power law at low momenta, $k^{-s}$, with exponent $s=3/2$, which corresponds to a free turbulence period dominated by three-particle scatterings; this is verified by the comparative analysis of lattice simulations with wave kinetic theory, details of which are given in \cite{turb,therm}. Larger momenta is bounded by an ultra-violet cut-off \cite{turb,therm}, which we will model in the form of an exponential function, parameterised by $k_{0}$:
\begin{multline}
    \label{eq:10}
    f_{\chi}(k,t)=\left(\frac{t}{t_{0}}\right)^{-\frac{7}{10}}\left(\frac{k}{k_{0}}\left(\frac{t}{t_{0}}\right)^{-\frac{1}{5}}\right)^{-\frac{3}{2}}
    \times\\
    \exp\left[-\frac{k}{k_{0}}\left(\frac{t}{t_{0}}\right)^{-\frac{1}{5}}\right] .
\end{multline}

Our analysis proceeds in the perturbative reheating period, during the later stages of thermalisation, from $t\sim t_{0}$. Once the Hubble expansion rate has decreased to the order of the inflaton decay width,  the inflaton can efficiently transfer its energy into the SM via the Higgs portal.

\subsubsection{Reheating: Standard Model production}

The inflaton mass governs the dominant mechanism for reheating; as a result, we analyse the reheating period separately for two regions of the inflaton parameter space: for light inflaton, with $m_{\chi}<2m_{\text{h}}$; and for heavy inflaton, with $m_{\chi}>2m_{\text{h}}$.

The mechanism for SM production from light inflaton $(m_{\chi}<2m_{\text{h}})$ is dominated by the scattering process $\chi\chi\rightarrow hh/gg$, which has has been previously analysed in \cite{Anisimov:2008qs,Bezrukov:2009yw}. The light inflaton mass is bounded from below by ensuring quantum corrections to the inflaton's quartic self-coupling are sufficiently small, so that slow-roll inflation isn't ruined ($\alpha^{2}\lesssim0.1\beta$) \cite{Bezrukov:2009yw}:
\begin{align}
    m_{\chi}>100 \  \Bigg(\frac{\beta}{1.5 \times10^{-13}}\Bigg)^{\frac{1}{2}} \Bigg(\frac{10^{-7}}{\alpha}\Bigg)^{\frac{1}{2}}\text{ MeV}.
\end{align}
The mass is bounded from above by requiring the minimum reheating temperature to be greater than the sphaleron freeze-out temperature $(\sim150\GeV)$, to ensure efficient sphaleron conversion of lepton to baryon asymmetry \cite{Bezrukov:2009yw}:
\begin{align}
    m_{\chi}\lesssim 1.25 \  \Bigg(\frac{\beta}{1.5 \times10^{-13}}\Bigg)^{\frac{1}{2}}\GeV.
\end{align}

Here we analyse the reheating process for heavy inflaton $(m_{\chi}>2m_{\text{h}})$ in detail. In the restored electroweak symmetry regime, the dominant reheating mechanism is via the decay process $\chi\rightarrow hh/gg$, with decay width \cite{Anisimov:2008qs}: 
\begin{align}
    \label{eq:14}
    \Gamma_{\text{SM}}=\Gamma_{\chi\rightarrow hh}+\Gamma_{\chi\rightarrow gg}=\frac{\beta}{4\pi}\frac{m_{\text{h}}^{4}}{m_{\chi}^{3}} .
\end{align}
The upper mass bound is provided by ensuring the minimum reheating temperature is greater than the electroweak symmetry breaking scale, $T_{\text{EW}}=160\GeV$ \cite{tempEW}; the results of this are given later in the paper, where we evaluate the bound numerically.

\subsubsection{Dark matter production}

Neutrino Minimal Standard Model $(\nu \text{MSM})$ can explain the origin of baryon asymmetry and DM \cite{neutrinoMSM2} through the addition of three right-handed singlet neutrinos, or HNLs, $N_{I}$ $(I=1,2,3)$, to the SM. By extending $\nu$MSM with a $N_{I}-X$ Yukawa coupling \cite{Anisimov:2008qs,Shaposhnikov:2006xi}:
\begin{multline}
    \label{eq:16}
    L_{\nu MSM+X}= L_{\nu MSM}+\\
    \frac{1}{2}(\partial_{\mu}X)^{2}-\frac{f_{I}}{2}\bar{N_{I}}^{c}N_{I}X+h.c.+V(X,\Phi) ,
\end{multline}
spontaneous symmetry breaking in the inflaton sector, $X\rightarrow\langle X\rangle+\chi$, generates the sterile neutrino mass, 
\begin{align}
    \label{eq:13}
    m_{N}=\langle X\rangle f_{I} ,
\end{align} 
and a coupling to the inflaton field. As in the scalar sector, we assume that the massive parameters enter only in the inflaton operators, i.e.\ we do not add bare Majorana masses for $N_{I}$.
The active-sterile neutrino coupling is strongly constrained from above by the absence of X-rays observed from the radiative decay of sterile neutrinos ($f^{\nu}_{I\alpha}\lesssim10^{-15}$) \cite{Abazajian_2001,Adhikari_2017}; we can therefore assume $f^{\nu}_{I\alpha}\ll f_{I}$, and neglect $N_{I}$ production from active-sterile neutrino oscillations. In this limit, all sterile neutrinos are produced via the freeze-in mechanism from inflaton decay, $\chi\rightarrow N_{I}\bar{N_{I}}$, in the early Universe. The corresponding decay width \cite{Bezrukov:2009yw,Shaposhnikov:2006xi},
\begin{align}
    \label{eq:17}
    \Gamma_{N}=\frac{f_{I}^{2}m_{\chi}}{16\pi} ,
\end{align}
is much less than the Hubble expansion rate throughout reheating, so sterile neutrinos remain decoupled from the thermal bath. The lightest of the three sterile neutrinos, $N_{1}$, is both massive and stable, and so is an ideal Feebly Interacting Massive Particle (FIMP) DM candidate. We will assume in our analysis that $N_{1}$ makes up the total DM energy density in the Universe, thereby constraining its abundance using \cite{PhysRevD}:
\begin{align}
    \label{eq:18}
    \Omega^{0}_\text{{DM}}&=\frac{s_{0}}{\rho_{\text{c}}}m_{N}Y_{N}(t)\sim 0.25 .
\end{align}
Here $s_{0}$ is the entropy density of the Universe today; $\rho_{\text{c}}$ is the critical density; and $Y_{N}$ is the ratio of the sterile neutrino number density $n_{N}$, over entropy, $s(t)$, evaluated once all DM has been produced. Constraining the coupling $f_{1}$ from the abundance, we can then evaluate the mass of DM, using (\ref{eq:13}), which has important consequences for structure formation \cite{Viel_2008}.

\section{Boltzmann equations for reheating and DM production}

Three Boltzmann collision integral equations fully describe the dynamics of the system of particles during reheating \cite{Kawasaki:1992kg,Kang:1993xz}:
\begin{widetext} 
\begin{align}
    \label{eq:21}
    \hat{L}f_{\chi}&=C^{\chi}_{\chi\leftrightarrow hh/gg}+C_{\chi\chi\leftrightarrow hh/gg}+C^{\chi}_{\chi\leftrightarrow N\bar{N}} 
    \\ \newline
    \label{eq:22}
    \hat{L}f_{N}&=C^{N}_{\chi\leftrightarrow N\bar{N}} 
    \\ \newline
    \label{eq:23}
    \hat{L}f_{SM}&=C^{\text{SM}}_{\chi\leftrightarrow hh/gg}+C^{\text{SM}}_{hh/gg\leftrightarrow \text{SMSM}}+C^{\text{SM}}_{h/g\leftrightarrow \text{SMSM}} 
\end{align}
The rate of particles $x$ scattering in and out of their distribution functions, $f_{x}$, is calculated using the collision integral; the standard definition for a general $2-2$ scattering process, $a(p)a'(p')\leftrightarrow b(q)b(q')$, with amplitude $M_{aa'\rightarrow bb'}$ is
\begin{align}
    \label{eq:24}
    C^{a}_{aa'\leftrightarrow bb'}&=\frac{1}{2E^{a}_{p}}\int\frac{g_{a'} \ d^{3}p'}{2E^{a'}_{p'}(2\pi)^{3}}\frac{g_{b} \ d^{3}q}{2E^{b}_{q}(2\pi)^{3}}\frac{g_{b'} \ d^{3}q'}{2E^{b'}_{q'}(2\pi)^{3}}
    \\ \nonumber
    &\ \ \ \ (2\pi)^{4}\delta^{4}(p+p'-q-q')|M_{aa'\rightarrow bb'}|^{2}\bigg[f_{b}(q,t)f_{b'}(q',t)-f_{a}(p,t)f_{a'}(p',t)\bigg] .
\end{align}
\end{widetext}
$E^{x}_{p}$ is the energy of particle $x$ with physical momemtum $p$, and $g_{x}$ is the number of effective degrees of freedom of particle $x$. 

Equations (\ref{eq:21}),(\ref{eq:22}) and (\ref{eq:23}) are simplified using the following:

\begin{itemize}
    \item $\Gamma_{\text{SM}}=O(10^{6}-10^{8})\Gamma_{N}$, so  $C^{\chi}_{\chi\leftrightarrow N\bar{N}}$ is neglected.
    \item $\Gamma_{\text{SM}}\gtrsim O(10^{7})\Gamma_{\chi\chi\rightarrow hh/gg}$, so $C_{\chi\chi\leftrightarrow hh/gg}$, is neglected.
    \item Sterile neutrino density remains low during reheating, as $\Gamma_{N}$ is much less than the Hubble expansion rate throughout the time of production; therefore the backward reaction $C^{N}_{N\bar{N}\rightarrow \chi}$ is neglected.
    \item Number conserving and violating interactions between $W^{\pm}/Z$ bosons are at rates of at least $O(10^{11})$ times greater than the Hubble expansion rate during reheating. As a result, kinetic and chemical equilibrium are reached very quickly, and so $C^{\text{SM}}_{hh/gg\leftrightarrow \text{SMSM}}$ and $C^{\text{SM}}_{h/g\leftrightarrow \text{SMSM}}$ are neglected. Additionally, we assume the SM thermalises instantaneously on production to some temperature $T_{\mathrm{SM}}$, and therefore implement the detailed balance condition.
\end{itemize}

The above simplifications reduces the set of Boltzmann equations (\ref{eq:21}, \ref{eq:22}, \ref{eq:23}) to the following \cite{steriledm,Merle_2015}:
\begin{align}
    \label{eq:30}
    \frac{\partial f_{\chi}(k,t)}{\partial t}&=\frac{am_{\chi}}{\sqrt{(am_{\chi})^{2}+k^{2}}}\Gamma_{\text{SM}}\Big[f^{eq}_{\chi}(T_{\text{SM}})-f_{\chi}(k,t)\Big] ,
    \\ 
    \label{eq:31}
    \frac{\partial f_{N}(k_{N},t)}{\partial t}&=\frac{m_{\chi}\Gamma_{N}a}{k_{N}^{2}}\int^{\infty}_{k'_{min}}dk'\frac{k'}{\sqrt{(am_{\chi})^{2}+k'^{2}}}f_{\chi}(k',t) .
\end{align}
$k$ and $k_{N}$ are the conformal inflaton and sterile neutrino momenta respectively. The lower bound on the integral is
\begin{align}
    k'_{\text{min}}=\Big|k_{N}-\frac{\big(am_{\chi}\big)^{2}}{4k_{N}}\Big|.
\end{align}
$f^{\text{eq}}_{\chi}(k,t)$ is the Bose-Einstein distribution function of the inflaton thermalised at the SM temperature, $T_{\text{SM}}(t)$:
\begin{align}
    \label{eq:27}
    f^{eq}_{\chi}(k,t)=\frac{1}{(2\pi)^{3}}\frac{1}{\text{exp}\Big[\frac{\sqrt{(a m_{\chi})^{2}+k^{2}}}{a T_{\text{SM}}(t)}\Big]-1} .
\end{align}

The differential equation for $a(t)$ is found using the Friedmann equation for the Hubble expansion rate, neglecting the contribution of the sterile neutrinos to the total energy density:
\begin{align}
    \label{eq:36}
    H(t)=\frac{\dot{a}(t)}{a(t)}=\sqrt{\frac{\rho_{\chi}(t)+\rho_{\text{SM}}(t)}{3M_P^{2}}} ;
\end{align}
where $\dot{a}(t)=\frac{da}{dt}$, $M_{\text{P}}$ is the reduced Planck mass and the energy densities of the inflaton and SM are:
\begin{align}
    \label{eq:34}
    \rho_{\chi}(t)=\frac{4\pi g_{\chi}}{a ^{4}} \int_{k=0}^{\infty}&dk \ k^{2}\sqrt{k^{2}+(a m_{\chi})^{2}}f_{\chi}(k,t) ,
    \\
     \label{eq:35}
    \rho_{\text{SM}}(t)&=\frac{\pi^{2}g_{\text{SM}}}{30}T_{\text{SM}}^{4}(t) .
\end{align}
The number of degrees of freedom of the inflaton is $g_{\chi}=1$. We take the number of effective degrees of freedom of the SM as constant throughout reheating, at $g_{\text{SM}}=100$. This is an acceptable approximation even for lighter inflaton, which produce sterile neutrinos at $(20<T_{\text{SM}}<80)\GeV$; with $g_{\text{SM}}=86.25$ \cite{dof}, the corresponding error on the inflaton-sterile neutrino coupling, $f_{1}$, is still less than $5\%$.

The differential equation for $T_{\text{SM}}$ is derived from the covariant conservation of the energy-momentum tensor:
\begin{multline}
    \label{eq:40}
    \frac{dT_{\text{SM}}}{dt}=-H T_{\text{SM}}-
    \\
    \frac{30}{g_{\text{SM}}\pi(aT_{\text{SM}})^{3}}\int^{\infty}_{0}dk \ k^{2}m_{\chi}\Gamma_{\text{SM}}\Big[f_{\chi}^{eq}(T_{\text{SM}})-f_{\chi}(k,t)\Big],
\end{multline}
and replaces the Boltzmann equation \eqref{eq:23}.

(\ref{eq:30}), (\ref{eq:36}) and (\ref{eq:40}) form a closed set of differential equations; the solution for $f_{\chi}(k,t)$ will then be used in (\ref{eq:31}) to solve for $f_{N}(k_{N},t)$.

\section{Analytical treatment}

The DM mass and momentum heavily depends on the inflaton distribution function at the time of production. At the extremities of the inflaton parameter space, where the reheating temperature is much less/greater than the inflaton mass, the DM is produced via two different mechanisms; here we define the reheating temperature, $T_{\text{eq}}$, as the temperature when the energy densities of the inflaton and the SM are equal. In these two regions we can use analytical approximations of the inflaton distribution functions to investigate the dependence of the DM properties on the inflaton parameters. In the parameter space between, the DM is produced via both mechanisms, so a more careful numerical analysis is required.

\subsection{\texorpdfstring{Relativistic inflaton particles $(T_{\text{eq}}\gg m_{\chi})$}{Relativistic inflaton particles (Teq>>mchi)}}

Light inflaton with larger self-couplings have $T_{\text{eq}}\gg m_{\chi}$, as shown in Figure \ref{fig:1}. In this limit, the inflaton has thermalised with the SM prior to the production of sterile neutrinos, at $T_{\text{SM}}\sim \frac{m_{\chi}}{2}$. The thermal inflaton distribution function at the SM temperature, given by (\ref{eq:27}), is used in (\ref{eq:31}) to obtain the analytical approximation of the sterile neutrino distribution. The sterile neutrino number density is given by \cite{Shaposhnikov:2006xi}:
\begin{align}
\label{eq:19}
    n_{N}(t)&=\frac{4\pi g_{N}}{a ^{3}} \int_{k=0}^{\infty}dk_{N} \ k_{N}^{2}f_{N}(k_{N},t)
    \\ \nonumber
    \label{eq:50}
    &=\frac{3\Gamma_{N}M_{0}\zeta(5)T_{\text{SM}}^{3}(t)}{2\pi m_{\chi}^{2}};
\end{align}
where the number of degrees of freedom of the sterile neutrino $g_{N}=2$, and $M_{0}\approx\frac{3M_{\text{P}}}{\sqrt{g_{\text{SM}}}}$. Having evaluated the entropy of the Universe at the end of reheating, 
\begin{align}
    s(t)=\frac{4}{3}\frac{\rho_{\text{SM}}(t)}{T_{\text{SM}}(t)}&=\frac{2\pi^{2}}{45}g_{\text{SM}}T_{\text{SM}}^{3}(t),
\end{align}
the relative DM abundance (\ref{eq:18}), is calculated using  (\ref{eq:2}), (\ref{eq:13}) and (\ref{eq:17}) for $\langle X\rangle$, $m_{N}$ and $\Gamma_{N}$ respectively, to obtain the following power-law relations between $m_{\chi}$, $\beta$, $f_{1}$ and $m_{N}$:
\begin{align}
\label{eq:51}
    \Omega^{0}_{\text{DM}}\propto& \frac{f_{1}^{3}}{\sqrt{\beta}},
    \\
    \label{eq:60}
    m_{N}\sim(9.94\times&10^{-6}) \ \frac{m_{\chi}}{\beta^{\frac{1}{3}}}\text{ MeV} .
\end{align}

The analytical approximation of the average momentum over temperature at the end of reheating is \cite{Shaposhnikov:2006xi}:
\begin{align}
\label{eq:52}
    \frac{\langle p_{N}\rangle}{T}=2.45 .
\end{align}

\subsection{\texorpdfstring{Non-relativistic inflaton particles $(T_{\text{eq}}\ll m_{\chi})$}{Non-relativistic inflaton particles (Teq << mchi)}}

Heavier inflaton with smaller self-couplings has  $T_{\text{eq}}\ll m_{\chi}$, as shown in Figure 1. In this limit, sterile neutrinos are produced prior to thermalisation from a highly infra-red inflaton distribution, at the same time as SM production:
\begin{align}
\label{eq:53}
    f_{\chi}(k,t)&\sim{e}^{-\Gamma_{\text{SM}}t}f_{\chi}(k,0)=\Big(\frac{k}{k_{0}}\Big)^{-\frac{3}{2}}e^{-\big(\frac{k}{k_{0}}+\Gamma_{\text{SM}}t\big)} .
\end{align}
The sterile neutrino number density is obtained by solving the equation
\begin{align}
\label{eq:54}
    \frac{\partial n_{N}(t)a^{3}}{\partial t}&=2\Gamma_{N}n_{\chi}(t)a^{3},
\end{align}
where the inflaton number density is
\begin{equation}
\label{eq:55}
    n_{\chi}(t)=\frac{4\pi g_{\chi}}{a^{3}} \int_{0}^{\infty}\!\!dk \,k^{2}f_{\chi}(k,t)
    =(2\sqrt{\pi})^{\frac{3}{2}}\left(\frac{k_{0}}{a}\right)^{3} {e}^{-\Gamma_{\text{SM}}t} .
\end{equation}
This leads to
\begin{equation}
\label{eq:56}
    n_{N}(t)=2(2\sqrt{\pi})^{\frac{3}{2}}\frac{\Gamma_{N}}{\Gamma_{\text{SM}}}  \Bigg(\frac{k_{0}}{a }\Bigg)^{3} .
\end{equation}
Solving for the entropy in terms of $k_{0}$:
\begin{align}
\label{eq:57}
    \frac{\partial n_{SM}(t)a^{3}}{\partial t}&=2\Gamma_{\text{SM}}n_{\chi}(t)a^{3} ,
    \\ \nonumber
    s(t)=\frac{2\pi^{4}n_{SM}(t)}{45\zeta(3)}&=\frac{4\pi^{4}(2\sqrt{\pi})^{\frac{3}{2}}}{45\zeta(3)} \Bigg(\frac{k_{0}}{a }\Bigg)^{3} ,
    \\ \nonumber
    Y_{N}&=\frac{45\zeta(3)}{4\pi^{4}} \frac{\Gamma_{N}}{\Gamma_{\text{SM}}} .
\end{align}
The relative DM abundance, (\ref{eq:18}), is calculated using (\ref{eq:2}), (\ref{eq:13}) and (\ref{eq:17}) for $\langle X\rangle$, $m_{N}$ and $\Gamma_{N}$, to obtain the following power-law relations between $m_{\chi}$, $\beta$, $f_{1}$ and $m_{N}$:
\begin{align}
\label{eq:58}
    \Omega^{0}_{\text{DM}}&\propto m_{N} \frac{\Gamma_{N}}{\Gamma_{\text{SM}}} \propto \frac{f_{I}^{3}m_{\chi}^{5}}{\beta^{\frac{3}{2}}},
     \\ 
\label{eq:59}
     m_{N}\sim&(1.05\times 10^{3}) \ m_{\chi}^{-\frac{2}{3}}\text{ MeV}.
\end{align}
Given that sterile neutrinos are produced at $T_{\text{SM}}\sim T_{\text{eq}}$, and presuming all the sterile neutrinos are created from inflaton particles at rest, their average momentum at the end of reheating is:
\begin{align}
\label{eq:67}
    \frac{\langle p_{N}\rangle}{T}\sim\frac{m_{\chi}}{2T_{\text{eq}}} .
\end{align}

\section{Numerical results and discussion}

We assume the free turbulent evolution of the inflatons driven by three particle scatterings up to some moment $t=t_0$, leading to the distribution function (\ref{eq:10}).
Starting from a Universe filled with only inflaton particles at $t_{0}$, the initial Hubble expansion rate is approximated by 
\begin{align}
    \label{eq:43}
    H(t_{0})=\sqrt{\frac{\rho_{\chi}(t_{0})}{3M_{\text{P}}^{2}}}
    \sim\sqrt{\frac{m_{\chi} k_{0}^{3}}{3M_{\text{P}}^{2}a(t_{0})^{3}}} ;
\end{align}
the parameter $k_{0}$ is chosen so that $H(t_{0})\gtrsim\Gamma_{\text{SM}}$, i.e. we choose the moment slightly before the inflaton decays. When the Hubble expansion rate has decreased to $H(t)\sim \Gamma_{\text{SM}}$, the Universe will start to reheat. 

\begin{figure}
  \includegraphics[width=\linewidth]{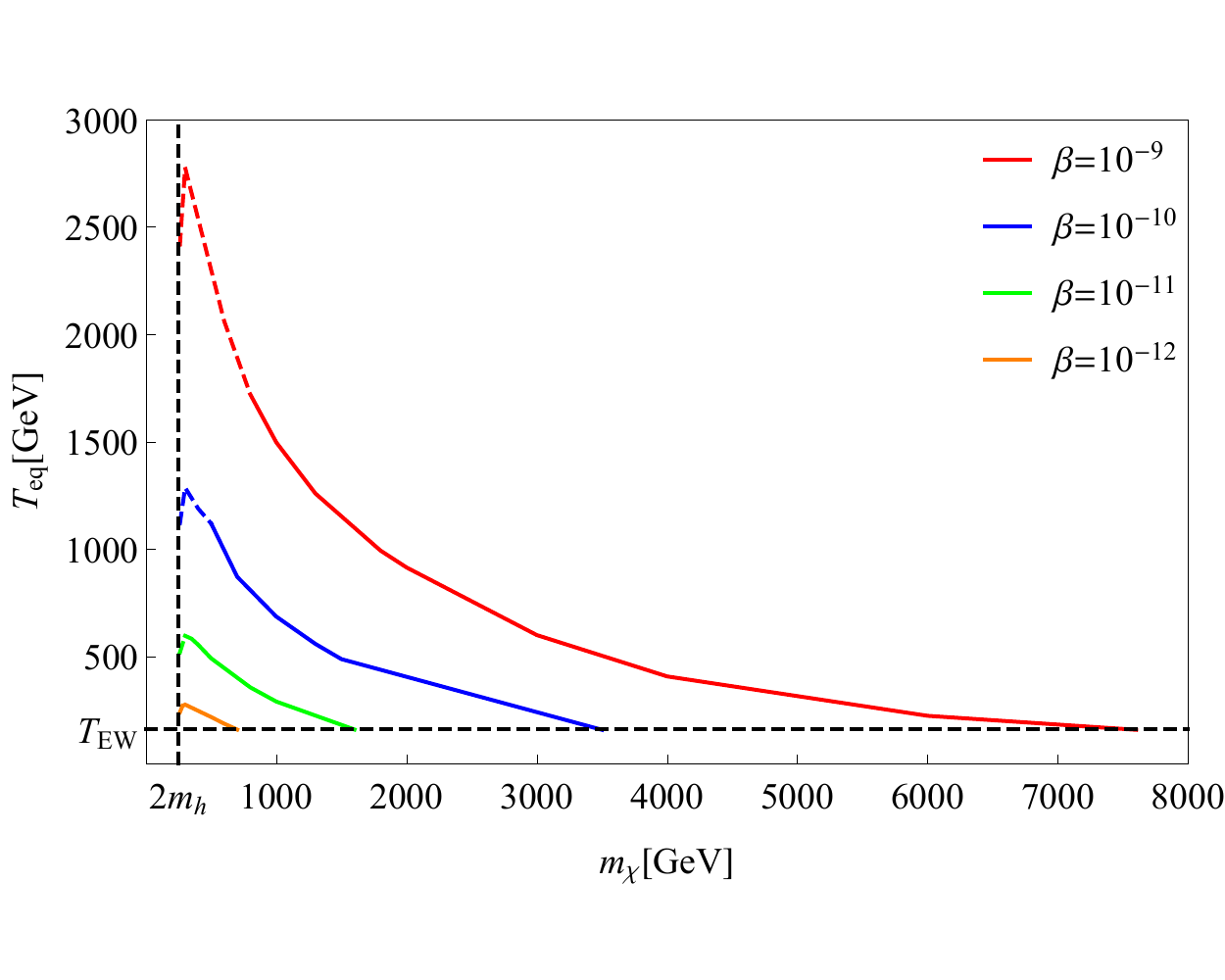}
  \vspace*{-15mm} 
  \caption{Reheating temperature, $T_{\text{eq}}$, defined when the energy densities of the SM and inflaton are equal, against the inflaton mass. The dotted lines give the lower inflaton mass bound, at $2m_{\text{h}}=250$ GeV, and the lower reheating temperature bound at the electroweak symmetry breaking scale, $T_{\text{EW}}=160$ GeV \cite{tempEW}.\label{fig:1}}
\end{figure}

First we need to define the parameter space of the heavy inflaton with self-couplings in the range $O(10^{-13})\leqslant\beta\leqslant O(10^{-9})$. The inflaton mass is bounded from below by the kinematics of the decay, $\chi\rightarrow hh/gg$, requiring $m_{\chi}>2m_{\text{h}}$. The inflaton mass is bounded from above by the minimum reheating temperature, which is required to exceed the electroweak symmetry breaking scale, $T_{\text{EW}}=160\GeV$ \cite{tempEW}. For inflaton that entered thermal equilibrium, we can roughly estimate
\begin{align}
    T_{\text{eq}}\propto\sqrt{\Gamma_{\text{SM}}}\propto\sqrt{\frac{\beta}{m_{\chi}^{3}}} .
\end{align}
However, the proper equilibrium temperature for inflaton starting from a non-thermal distribution is found by solving the Boltzmann equations numerically. The mass bounds for inflaton with couplings $(10^{-12}\leqslant\beta\leqslant10^{-9})$ are defined in Figure \ref{fig:1}; inflaton with coupling $\beta=10^{-13}$ is not included here, as $T_{\text{eq}}<T_{\text{EW}}$ for $m_{\chi}>2m_{\text{h}}$. 

The dotted lines in Figure \ref{fig:1} give the area of parameter space where we would expect the thermal mass of the Higgs boson to suppress the inflaton decay width. Precise analysis of this region requires full thermal quantum treatment of the evolution, which is beyond the scope of this article, so the precise results at $m_\chi\sim2m_h$ should be treated with caution.

\begin{figure}
  \includegraphics[width=\linewidth]{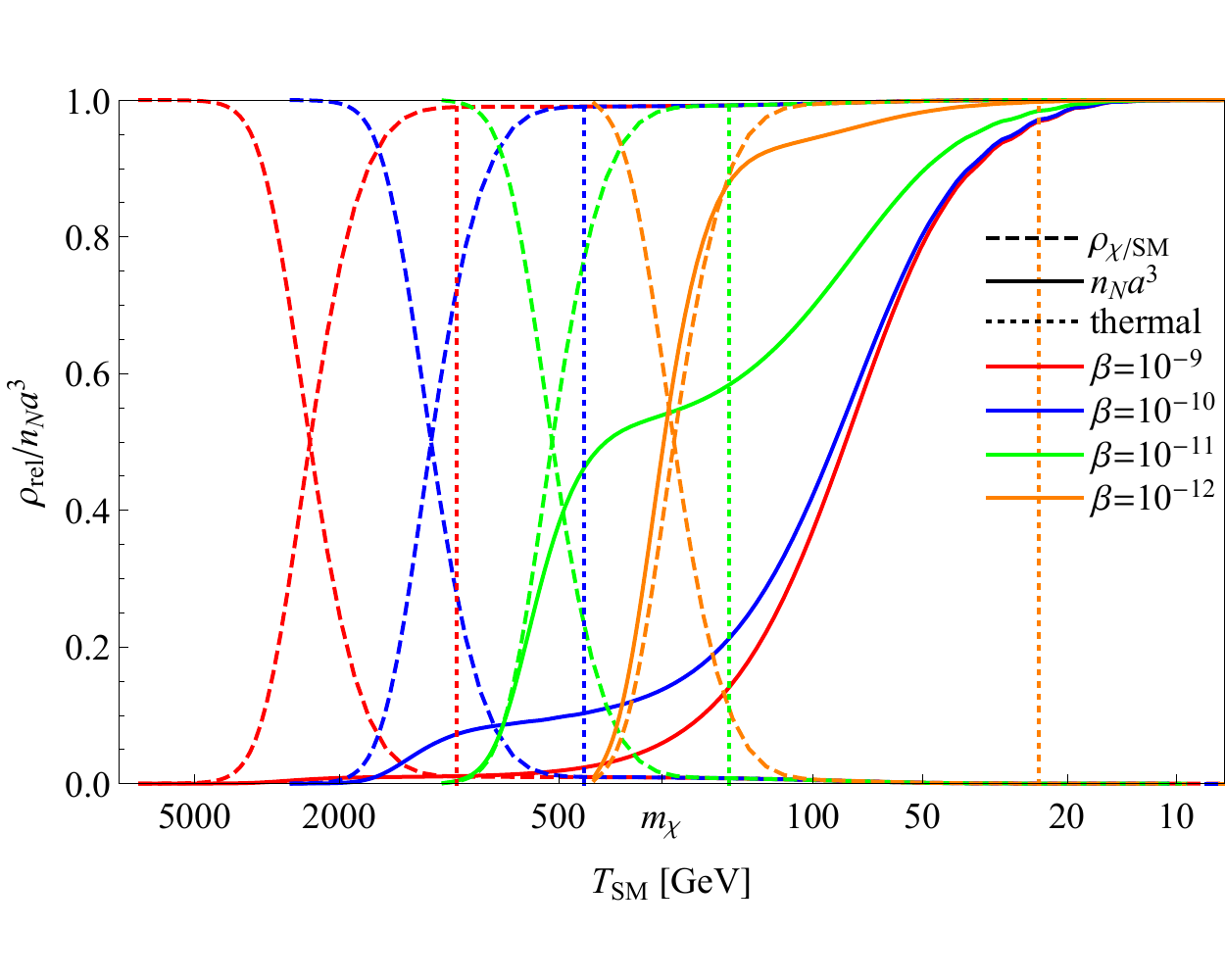}
  \vspace*{-12mm} 
  \caption{The plot gives numerical results from $260\GeV$ inflaton, with self-coupling $(10^{-12}\leqslant\beta\leqslant10^{-9})$, represented by different colours given in the legend. The dashed lines are the relative energy densities of inflaton/SM, $\rho_{\chi}/\rho_{\text{SM}}$, and the full lines are the normalised conformal number densities of sterile neutrino, $n_{N}a^{3}$, plotted against the SM temperature, $T_{\text{SM}}$. The vertical dotted lines give the SM temperature at which the inflaton distribution thermalises.
  \label{fig:2}}
\end{figure}

The reheating temperature relative to the inflaton mass determines when the DM leptons are produced relative to the thermalisation of the inflaton distribution with the SM, and therefore the properties of the DM. This is demonstrated in Figure \ref{fig:2}, which shows the time of sterile neutrino and SM production from $260 \GeV$ inflaton, as a function of the SM temperature. The temperature when the inflaton distribution thermalises is indicated by the vertical dotted lines. 

Inflaton particles of mass $260 \GeV$ and coupling $\beta=10^{-9}$ have a reheating temperature of $T_{\text{eq}}\sim 2400\GeV$. As shown by the red line in Figure \ref{fig:2}, sterile neutrinos are produced from remnant thermalised inflaton particles post SM production. This is most efficient when the Universe has cooled to $T_{\text{SM}}\sim m_{\chi}/2$; at lower temperatures, production is inefficient as the inflaton occupation number is highly Boltzmann suppressed. However, non-relativistic inflaton most efficiently produce sterile neutrinos at the same time as the SM, at $T_{\text{SM}}\sim T_{\text{eq}}$; at this time, the inflaton distribution is non-thermal and the occupation number is at its largest. The orange line in Figure \ref{fig:2} demonstrates the non-thermal  production of sterile neutrinos from inflaton particles of mass $260 \GeV$, coupling $\beta=10^{-12}$, and a reheating temperature of $T_{\text{eq}}\sim 240\GeV$. The intermediate couplings, $\beta=10^{-10}/10^{-11}$, generate sterile neutrinos by both mechanisms that govern the highly relativistic/non-relativistic inflaton regions. The blue and green lines in Figure 2 show the increasing efficiency of sterile neutrinos production at $T_{\text{eq}}$ with decreasing $\beta$.

\begin{figure}
    \includegraphics[width=0.93\linewidth]{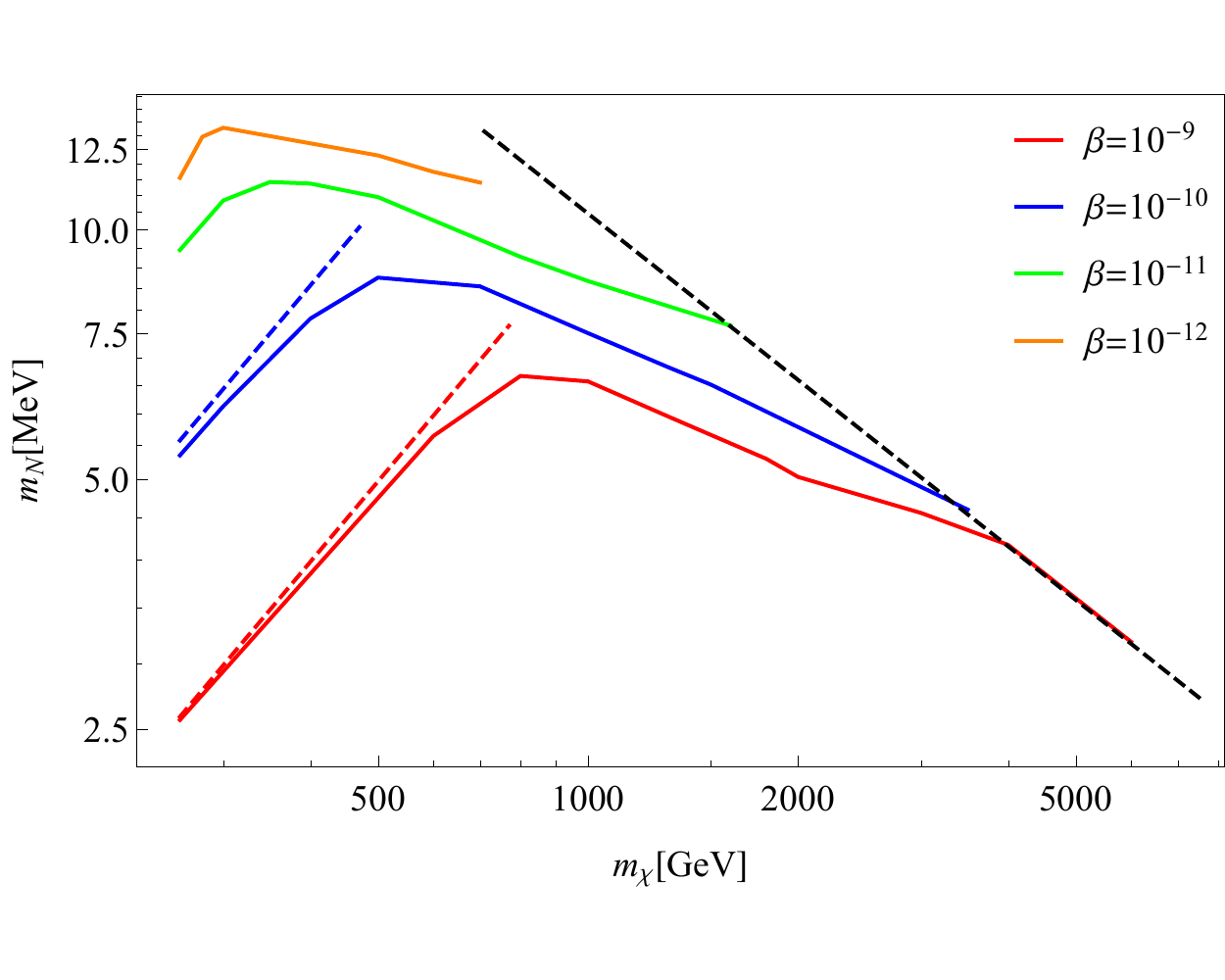}\\[-10mm]
    \includegraphics[width=1.07\linewidth]{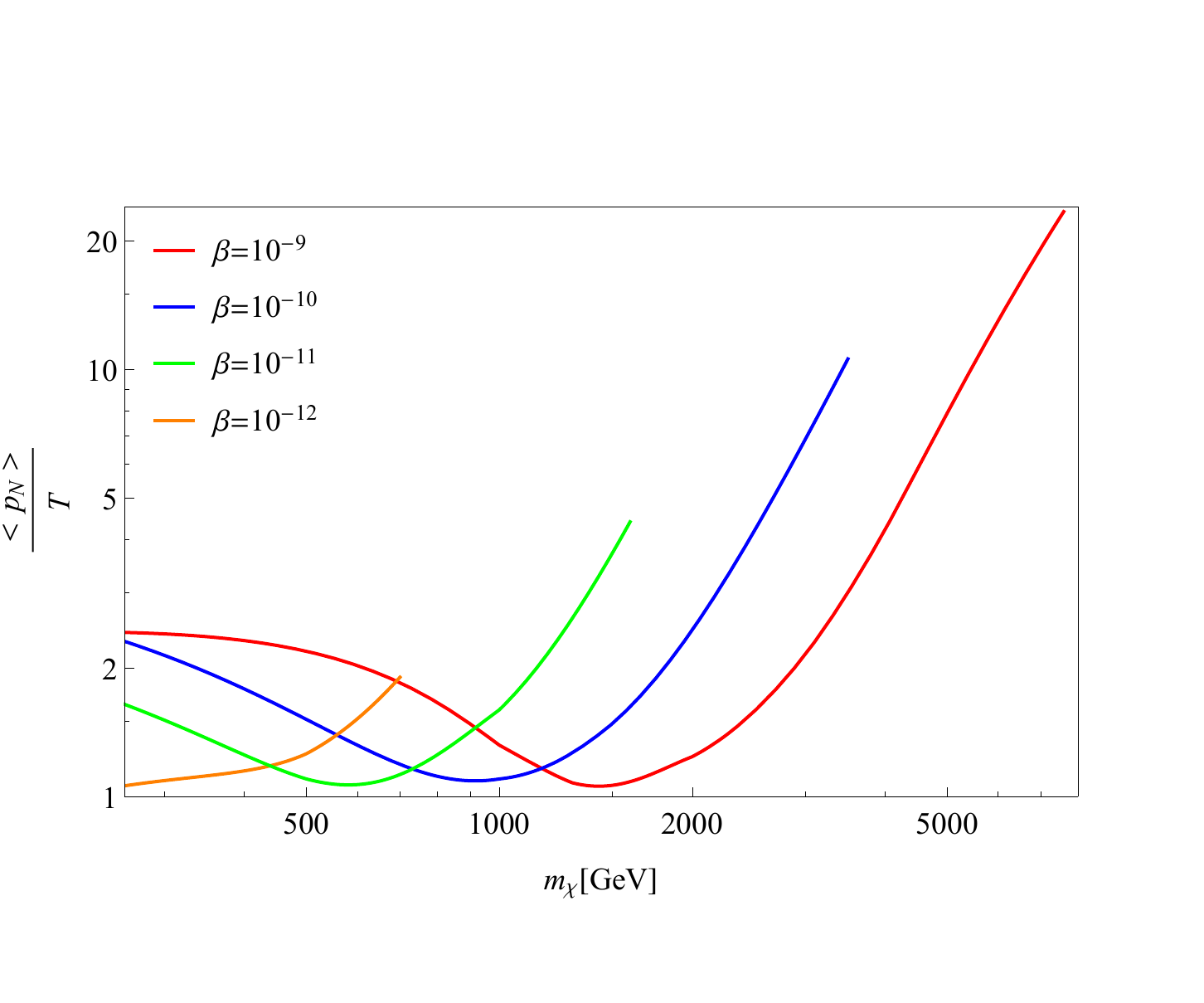}
     \caption{(top) Sterile neutrino mass against inflaton mass. The solid lines are the numerical results and the dashed lines are the analytical approximations. The analytical results for thermal inflaton, given by (\ref{eq:60}), have a dependence on $\beta$ so are colour-coded accordingly. The analytical result for non-relativistic inflaton, given by (\ref{eq:59}), has no dependence on $\beta$, so is given by the black dashed line.
     (bottom) Average sterile neutrino momentum over temperature at the end of reheating, $\langle p_{N}\rangle/T$, against inflaton mass. Different colours represent inflaton with different self-coupling, $\beta$.
     \label{fig:3}}
\end{figure}

Analytical approximations of the sterile neutrino mass as a function of inflaton mass, that lead to the proper DM abundance,  show a positive correlation for relativistic inflaton particles (\ref{eq:60}), and a negative correlation for non-relativistic inflaton particles (\ref{eq:59}). Plotting the sterile neutrino mass against the inflaton mass, shown in Figure \ref{fig:3} (top), allows us to clearly identify which production mechanism dominates in different regions of the inflaton parameter space. The inflaton mass which produces the maximum sterile neutrino mass is analytically approximated, using (\ref{eq:60}) and (\ref{eq:59}), by:
\begin{align}
    \label{eq:65}
    m_{\chi}\sim (5.1 \times 10^{4})\beta^{\frac{1}{5}} \GeV;
\end{align}
the left/right of the peak corresponds to the inflaton parameter space where the thermal/non-thermal production mechanism dominates. In agreement with Figure \ref{fig:3} (top), (\ref{eq:65}) demonstrates that as $\beta$ increases, the peak sterile neutrino mass moves to larger values of the inflaton mass; and (\ref{eq:60}) states that relativistic inflaton with smaller $\beta$ produce heavier sterile neutrinos.

The numerical results in Figure \ref{fig:3} (top) are given by the solid lines, and the analytical results by the dashed lines. The relativistic approximations (\ref{eq:60}), have a dependence on $\beta$, and are coloured accordingly; the non-relativistic approximation (\ref{eq:59}), has no dependence on $\beta$, so is given by the black line. The analytical and numerical results match to good accuracy for the lightest and heaviest inflaton particles with self-coupling $\beta=10^{-9}$. The inflaton particles with smaller $\beta$ do converge towards the analytical approximations, however numerical analysis is necessary for arbitrary values of the parameters.

Figure \ref{fig:3} (bottom) plots the average sterile neutrino momentum over temperature at the end of reheating, $\langle p_{N}\rangle/T$, across the inflaton parameter space. The lightest inflaton particles, with coupling $\beta=10^{-9}$, have $\langle p_{N}\rangle/T\sim2.4$, which is in agreement with our analytical approximation for thermal production (\ref{eq:52}). Whilst $m_{\chi}<T_{\text{eq}}$, increasing the inflaton mass increases the efficiency of sterile neutrino production at $T_{\text{eq}}$, thereby decreasing $\langle p_{N}\rangle/T$ until a minimum is reached at $T_{\text{eq}}=m_{\chi}$, corresponding to $\langle p_{N}\rangle/T \sim 1$. $\langle p_{N}\rangle/T$ rapidly increases once $m_{\chi}>T_{\text{eq}}$, as $\langle p\rangle$ and $T_{\text{eq}}$ are increasing and decreasing functions of $m_{\chi}$, respectively. Analytical results are consistent with our numerical results for heavy non-relativistic inflaton, given by (\ref{eq:67}); for example, the analytical estimate for $7600\GeV$ inflaton is $\langle p_{N}\rangle/T\sim24$.

Other models for sterile neutrino DM production include light scalar decay \cite{Shaposhnikov:2006xi}, the resonant or non-resonant production from active neutrinos \cite{DW1,DW2} and thermal production with further entropy dilution from the dark sector \cite{en1,en2}. These models produce keV sterile neutrinos; with an average momentum over temperature at active neutrino decoupling of $\langle p_{N}\rangle/T_{\nu}=O(1)$, they are warm DM candidates. By comparison, production in heavy inflaton decays needs MeV neutral leptons, which are Cold DM candidates with
\begin{align}
    \frac{\langle p_{N}\rangle}{T_{\nu}}=\bigg(\frac{g_{\text{SM}}(T_{\nu})}{g_{\text{SM}}(T)}\bigg)^{\frac{1}{3}}\frac{\langle p_{N}\rangle}{T}\sim0.5-11;
\end{align}
where $g_{\text{SM}}(T_{\nu})=10.75$. Our model is therefore well within the constraints from the Lyman-alpha data, which puts an upper bound on the DM free streaming parameter, or equivalently velocity $v<10^{-3}$ at temperature $\sim1\text{ eV}$ \cite{Petraki_2008,Viel_2008}. The velocity of the DM in our model is  
\begin{multline}
    \langle v\rangle_{T=1\eV}=10^{-6}\bigg(\frac{\text{ MeV}}{m_{N}}\bigg)\bigg(\frac{g_{\text{SM}}(1 \eV)}{g_{\text{SM}}(T)}\bigg)^{\frac{1}{3}}\bigg(\frac{\langle p_{N}\rangle}{T}\bigg)\\
    \sim O(10^{-8}-10^{-6}) , 
\end{multline}
where $g_{\text{SM}}(1\eV)\sim 3.91$ \cite{dof}. Sterile neutrinos with the highest velocity are produced from $7600\GeV$ inflaton with coupling $\beta=10^{-9}$, and the lowest velocity from inflaton with coupling $\beta=10^{-12}$.

\section{Conclusion}\label{sec:Conclusions}

We studied a singlet scalar model with a quartic self-interaction and a coupling to the Higgs sector. With the addition of a non-minimal coupling of the scalar field to gravity, this model can successfully produce inflation within CMB bounds of the tensor-to-scalar ratio and the amplitude of primordial scalar perturbations, for self-coupling in the range $\beta=O(10^{-13}-10^{-9})$. With scale invariance only broken in the scalar sector, the inflaton-Higgs coupling gives rise to symmetry breaking in the Higgs sector and provides the mechanism to initiate reheating. In the parameter space of heavy inflaton particles $(m_{\chi}>2m_{h})$, the mixing angle with the Higgs sector is very small, thus evading direct experimental constraints. Our analysis restricts the heavy inflaton mass range to $(250<m_{\chi}\lesssim7600)$ GeV, by ensuring efficient reheating of the Universe above the electroweak scale, via inflaton decay into two Higgs bosons.

A mechanism for freeze-in DM production is realised in our model through the addition of a Yukawa coupling of the inflaton to sterile neutrinos, within the framework of $\nu$MSM. We assume DM is made up entirely of the lightest sterile neutrino and is produced via inflaton decay. For inflaton with $m_{\chi}\ll T_{\text{eq}}$, DM is produced once the inflaton has thermalised and so the model parameters can be deduced analytically. For heavy inflaton with $m_{\chi}\gtrsim T_{\text{eq}}$, DM is produced simultaneously with the SM from a highly non-thermal infra-red inflaton distribution, and so it is necessary to solve the Boltzmann equations numerically. In the heavy inflaton parameter space the DM is strongly non-thermal and cold, with $\langle p_{N}\rangle/T\sim O(1-10)$ at the end of reheating. Using the known abundance of DM in the Universe, the Yukawa coupling constrains the DM mass to $O(1-10)\text{MeV}$, which puts our results well within the requirements for structure formation given by the Lyman-alpha data.

Let us turn to the limitations of our analysis. First, our results depend on the assumption that the initial inflaton distribution is governed by turbulence driven by 3-particle scatterings, as suggested in \cite{turb,therm}. We assessed the level of influence of these assumptions by comparing the results of the three-particle scattering function to a 4-particle scattering function (with the power law distribution $k^{-s}$ with $s=5/3$ instead of \eqref{eq:10}) in the non-analytical region of the parameter space. We found a relatively weak dependence on the initial distribution function, with up to a $(10-20)\%$ difference between the results.
Secondly, we ignored the details of symmetry restoration in the electroweak sector after preheating, which would require a full thermal field theory treatment. Thus our results for inflaton masses approaching the kinematic limit of decay into two Higgs bosons may be modified by exact study.

A potentially interesting region of inflaton masses could be when $m_\chi\simeq m_{\text{h}}$, when the mixing angle \eqref{eq:5} becomes large. However, we expect that reheating in this range is still inefficient. Although this would significantly enhance the inflaton decay rate via inflaton-Higgs mixing, such processes can not contribute to reheating in restored electroweak symmetry, and $\chi\chi\rightarrow hh$ is inefficient for inflaton in this mass range. Nonetheless, we can not rule out the possibility of significant SM production here as a result of the misalignment of the inflationary attractor with the vacuum, without a careful study of the preheating period. 

Overall, the model provides a viable inflationary mechanism and DM generation, while evading all current experimental constraints, due to extremely low mixing of the inflaton with the Higgs sector, and strongly sterile leptons at CDM velocity.

A future study that could lead to potentially interesting detectable signatures would require extensions of the basic model studied here. In particular, it is possible to modify the potential so that the DM is warmer and therefore more visible. Mass terms for the Higgs doublet and sterile neutrino were removed from the Lagrangian so scale invariance is only broken in the inflaton sector, however these terms can be used to tune the sterile neutrino mass so the DM is lighter and therefore warmer. For example, the Majorana sterile neutrino mass term, $-\frac{M_{I}}{2}\bar N^{c}N$, can be tuned to have the opposite sign and have an equal magnitude to that acquired from the Yukawa coupling, thereby giving a smaller effective sterile neutrino mass. Alternatively, the inclusion of the symmetry breaking Higgs doublet mass term, $+\mu^{2}H^{\dagger}H$, would change the VEV of the inflaton field, and thus change the contribution to the sterile neutrino mass from the Yukawa term. Producing lighter and therefore warmer DM would allow the model to be constrained from the observation of the smallest DM structures formed in the Universe.

Future work on this model may also include studying the effects of adding the renormalisable trilinear inflaton-Higgs coupling, $\chi H^{\dagger}H$. A small trilinear coupling is necessary to avoid a domain wall problem, however a sizeable coupling may significantly enhance $\chi\rightarrow hh$ in the heavy inflaton parameter space, thus reheating the Universe more efficiently and extending the upper mass bound of the inflaton. Additionally there are experimental motivations if $\theta_{\text{m}}$ is significantly larger, as new detection channels in particle colliders, such as $\chi\rightarrow q\bar{q}$, may become accessible.

\begin{acknowledgments}
The authors of the paper are grateful to D. Gorbunov for valuable discussions. The work is supported in part by STFC research grant ST/P000800/1.
\end{acknowledgments}

\bibliography{eprintcontrol,bibfile}

\end{document}